\documentclass[fleqn,12pt]{wlscirep}
\usepackage{wrapfig, framed, caption}
\usepackage[utf8]{inputenc}
\usepackage{heuristica}
\usepackage{graphicx}
\usepackage{wrapfig, framed, caption}
\usepackage{tcolorbox}
\usepackage{xr}
\usepackage{hyperref}
\usepackage{xcolor}

\makeatletter
\newcommand*{\addFileDependency}[1]{
  \typeout{(#1)}
  \@addtofilelist{#1}
  \IfFileExists{#1}{}{\typeout{No file #1.}}
}
\makeatother

\usepackage{setspace}
\usepackage[normalem]{ulem}
\useunder{\uline}{\ulined}{}
\usepackage{xparse}
\newsavebox{\fminipagebox}
\NewDocumentEnvironment{fminipage}{m O{\fboxsep}}
 {\par\kern#2\noindent\begin{lrbox}{\fminipagebox}
  \begin{minipage}{#1}\ignorespaces
 \end{minipage}\end{lrbox}%
  \makebox[#1]{%
    \kern\dimexpr-\fboxsep-\fboxrule\relax
    \fbox{\usebox{\fminipagebox}}%
    \kern\dimexpr-\fboxsep-\fboxrule\relax
  }\par\kern#2
 }

\title{A Network Map of {\it The Witcher}}

\author[*1,2,3]{Mil\'an Janosov}

\affil[1]{Department of Network and Data Science, Central European University, Budapest, 1051, Hungary}

\affil[2]{Datapolis Inc, Budapest, 1112, Hungary}

\affil[3]{Milan Janosov \href{https://linktr.ee/floraborsi}{https://linktr.ee/janosov}}

\affil[*]{janosovm@gmail.com}

\begin{document}

\maketitle


\section*{Abstract}
{\small

{\it The Witcher}, a fantasy novel series by Andrzej Sapkowski, has inspired multiple computer games and was even launched on Netflix in 2019, with the latest season airing in December 2021. Fictional storylines such as {\it The Witcher}, can be well summarized and simplified by network science, extracting the backbone of a complex series of events. In other words, network mapping and data science can transform thousands of pages into one simple, visual, social map. This exemplary visualization also highlights the power of network science by uncovering hidden patterns, such as influencers, hubs, and cliques in the interplay of hundreds of people and their countless encounters. As these methods can be applied to novels and their adaptations as well, they also allow us to compare the original and the adapted pieces, from the cast to the plots. Therefore, in this article, I provide a social network analysis about the societal characteristics of {\it The Witcher}, comparing the novels and the TV show.

}

\vspace{0.5cm}
{\small {\bf Keywords}: network science, fantasy novel, social network analysis}

\vspace{1.0cm}
{\it \hspace{-1cm} Published in Nightingale, Journal of the Data Visualization Society, December 23, 2021~\cite{nightingale}. }
\vspace{1.0cm}

\section{Popular interest}

\begin{figure}[!hbt]
\centering
\includegraphics[width=1.00\textwidth]{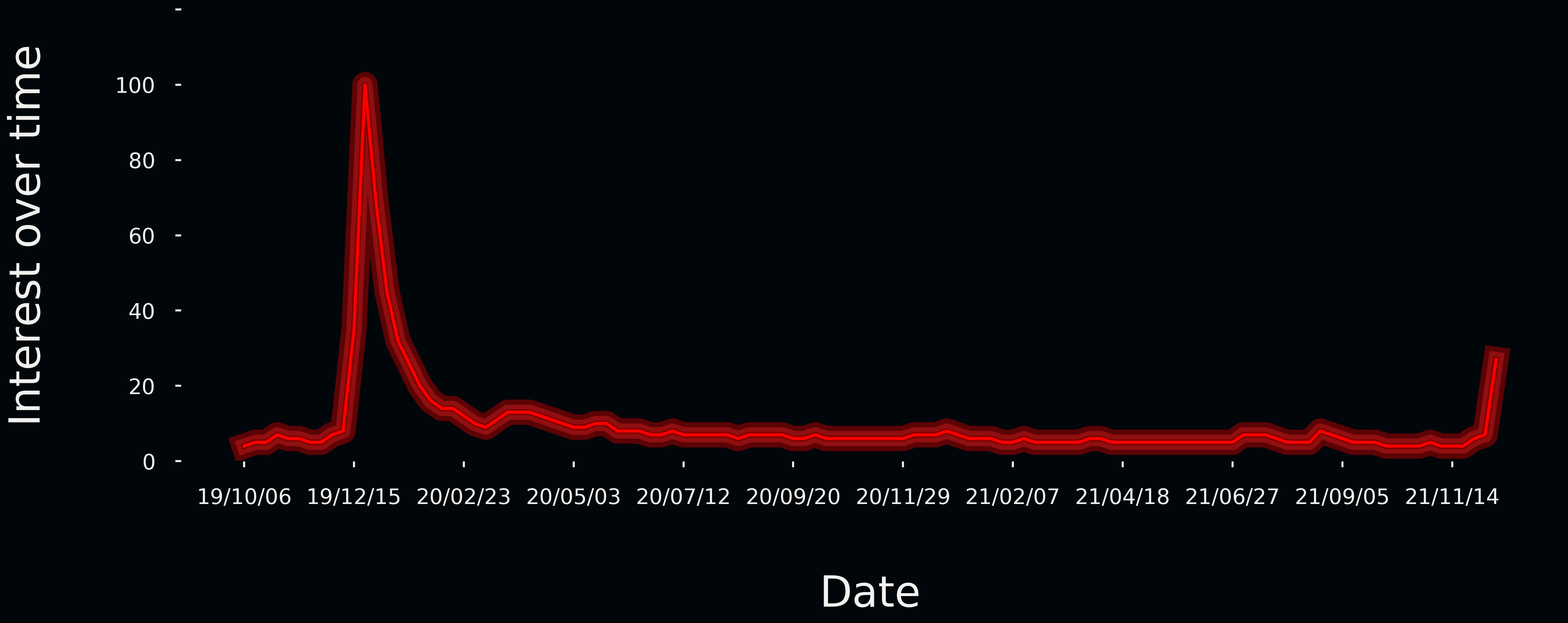}
\caption{The relative popularity of {\it The Witcher} over time, according to Google Trends.}
\label{fig:fig1}
\end{figure}

Using the public API of Google Trends, I can easily extract relative popularity using the search phrase ‘Witcher’ to infer the level of public interest in it. The evolution of popularity level, most likely driven by the TV show, is visualized in Figure \ref{fig:fig1}, with a pronounced peak in mid-December 2019, which overlaps with the release date (17 December 2019) of the first season of {\it The Witcher} on Netflix. While the series debuted on the week of the 15th, it only reached a relative popularity of 35 percent that week, while the 100 percent peak occurred a week later. Then, the first peak slowly started to decline, dipping below the popularity of the debut week six weeks later. About two years later, after the release of the second season, one week after the release, we already see a pronounced peak in interest, nearing the previous top.

\section{Data}

As a primary data source, I rely on the digitalized version of the following books from {\it The Witcher} series~\cite{books_w}: The Last Wish, The Sword of Destiny, Blood of Elves, Times of Contempt, Baptism of Fire, The Tower of the Swallow, and The Lady of the Lake – a total of about 2,500 pages, totaling more than one million words.

Additionally, I collected the list of characters in each book from {\it The Witcher} Wiki~\cite{wiki}, which lists 690 named characters across these books. The Sword of Destiny is the least populated book with 117 characters, while The Lady of the Lake features the biggest crowd of 200 individuals. Additionally, this platform named about a hundred characters appearing in the Netflix show as well~\cite{wiki2}.

\section{Network definition}

To extract the social network~\cite{got1, got2, asimov1} of {\it The Witcher}, I used the following text-processing procedure on each book. First, I tokenized every book into a list of sentences and then labeled each sentence by the name of the characters appearing in that sentence. Second, I defined a window size as the distance of two pairs of sentences and assumed that if two characters are mentioned in two sentences within this window then there is an edge with a weight of one between them. The following two quoted sentences illustrate when the two mentions occur in the same sentence or two sentences apart:

\begin{quote}
{\it \small  "Screw yourself!' screamed Yennefer, still trying to scratch Geralt's eyes out."} 

-- { \small The Last Wish}
\end{quote}

\begin{quote}
{\it \small "The sorceress courtesied once again, steadying herself firmly on Geralt's shoulder. Villentretenmerth stood up and looked at her, his face very serious. "Excuse my boldness and my frankness, Yennefer."}

-- {\small The Sword of Destiny}
\end{quote}

Then, I tested several window sizes (1, 3, 5, 6, 7), and based on the completeness of the network and the level of noise, I decided to use a window size of five sentences. Following this rule, I finally arrived at a network of 541 nodes (characters) among whom there are 3,306 links total, with a median connection strength of five.

\begin{figure}[!hbt]
\centering
\includegraphics[width=1.00\textwidth]{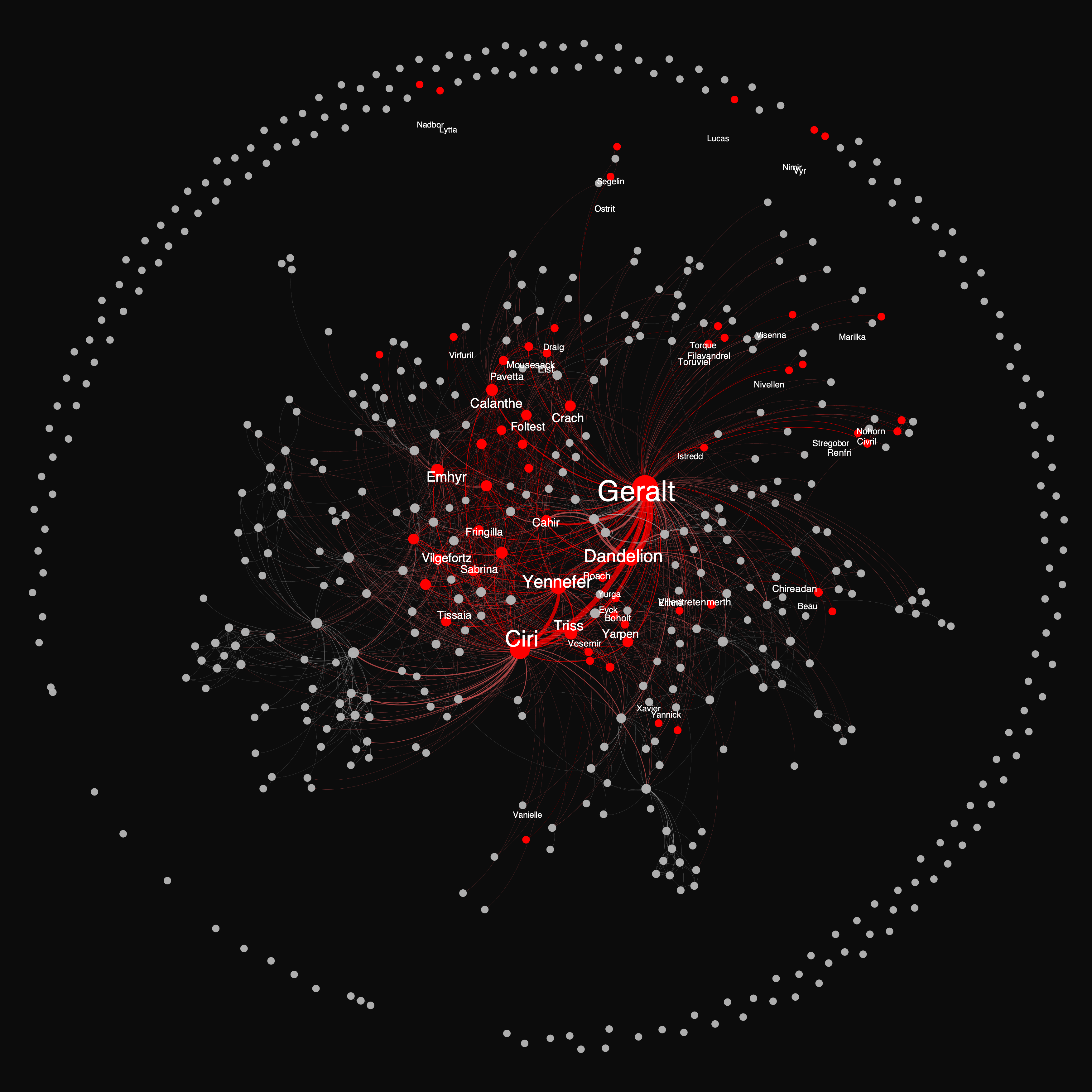}
\caption{The social map of {\it The Witcher}. Characters are represented by nodes, their size corresponding to their node degree (the number of connections), and their coloring and labeling show those individuals who have already appeared in the first season of {\it The Witcher}’s Netflix TV adaptation (red) from the rest of the characters yet to appear (grey). The network links are proportional to the number of times two characters were mentioned within a five-sentence distance from each other in the novel.}
\label{fig:fig2}
\end{figure}

Figure \ref{fig:fig2} visualizes this network according to a force atlas layout~\cite{force}, where the node sizes are proportional to the degree of their nodes, in other words, the number of connections they have~\cite{netsci}. While this network visualization already contains the full cast of the novel series, I only labeled and colored in red those individuals that have also appeared in the Netflix TV show as of the publication date of this article. This coloring already hints that there is quite some space for spoilers (next section), for instance, showing a great number of internal and external nodes uncolored, or entire clusters of grey nodes. In the next section, I conduct a more detailed analysis where I also compare this Netflix-overlap segment of the network to the full picture.

\section{Network characteristics}

\paragraph{The most important characters.} 

In social systems, some people seem to know everybody, some only have a few connections, and others might be limited to distinct groups. These different types of social status can be captured by network properties–mathematical measures called centrality measures. In this article I studied two of these measures. First, I computed the so-called node strength for each node: the sum of all the weights of an individual’s connections, weighting together both the quantity and the quality of one’s relationships. Then I split the analysis into two parts: comparing all the characters present in the novel to the ones that have only been in the TV show. Following this split, I created the top 20 list of most important characters, based on their node strength shown in Table \ref{tab:tab1}. When comparing these two rankings, I can immediately spot that the majority of the top characters in the novel have also appeared in the TV show. There are only a few other characters typically present in just one or two specific storylines, such as Bonhart (Ciri’s nemesis), and Geralt’s mission companions, Milva and Regis, who are still missing (marked by red) from Netflix, implying relatively few new major characters likely show up in the coming seasons.

\begin{table}
\centering
  \caption{The top list of individuals ranked based on their node strength. The first column shows only the characters appearing in the Netflix show, while the second one covers all the characters in the novels. The second column shows those characters appearing in the novel toplist, but not in the Netflix,  in red.}
  \label{tbl:excel-table}
  \includegraphics[scale=0.7]{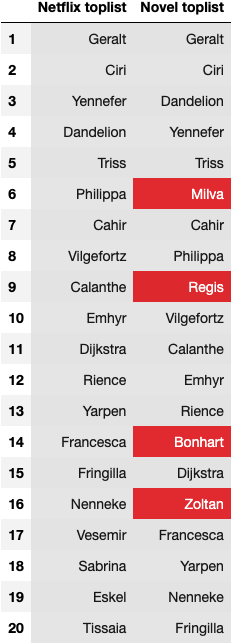}
  \label{tab:tab1}
\end{table}

Second, I computed a measure called betweenness centrality, which is a metric quantifying influence of a node over the flow of information in a network – typically, the bottlenecks or the bridges. While usually there is some correlation between a node’s betweenness and its strength, betweenness can also reveal nodes in a network that are barely visible, yet have a pronounced role in the connectivity of the network.

Table \ref{tab:tab2} ranks the top 10 highest betweenness individuals in the series and the books. For instance, you can see an expected high correlation in the case of Geralt, Yennefer, and Ciri as all are major characters. Keeping the Netflix filter on, you see other major characters from the screen. However, interestingly, there are a few lower strength nodes (not even in the top 20 of Table 1) in the top five list in the novels, such as: 

\begin{itemize}
    \item King Foltest, who may connect different parts of the realm as the ruler of a major kingdom, and 
    \item  his courtier, Ostrit, who is actually making these connections by venturing across the kingdoms, and finally, 
    \item  Yurga the traveling merchant, who helped out Geralt on his journey and aided in Geralt meeting Ciri (linking different storylines together).
\end{itemize}

\begin{table}[!hbt]
\centering
  \caption{The toplist of individuals based on their betweenness centrality, listing the most pronounced bridge roles in the show and the novels.}
  \label{tbl:excel-table}
  \includegraphics[scale=0.7]{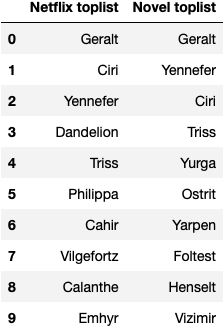}
  \label{tab:tab2}
\end{table}

\begin{figure}[!hbt]
\centering
\includegraphics[width=1.00\textwidth]{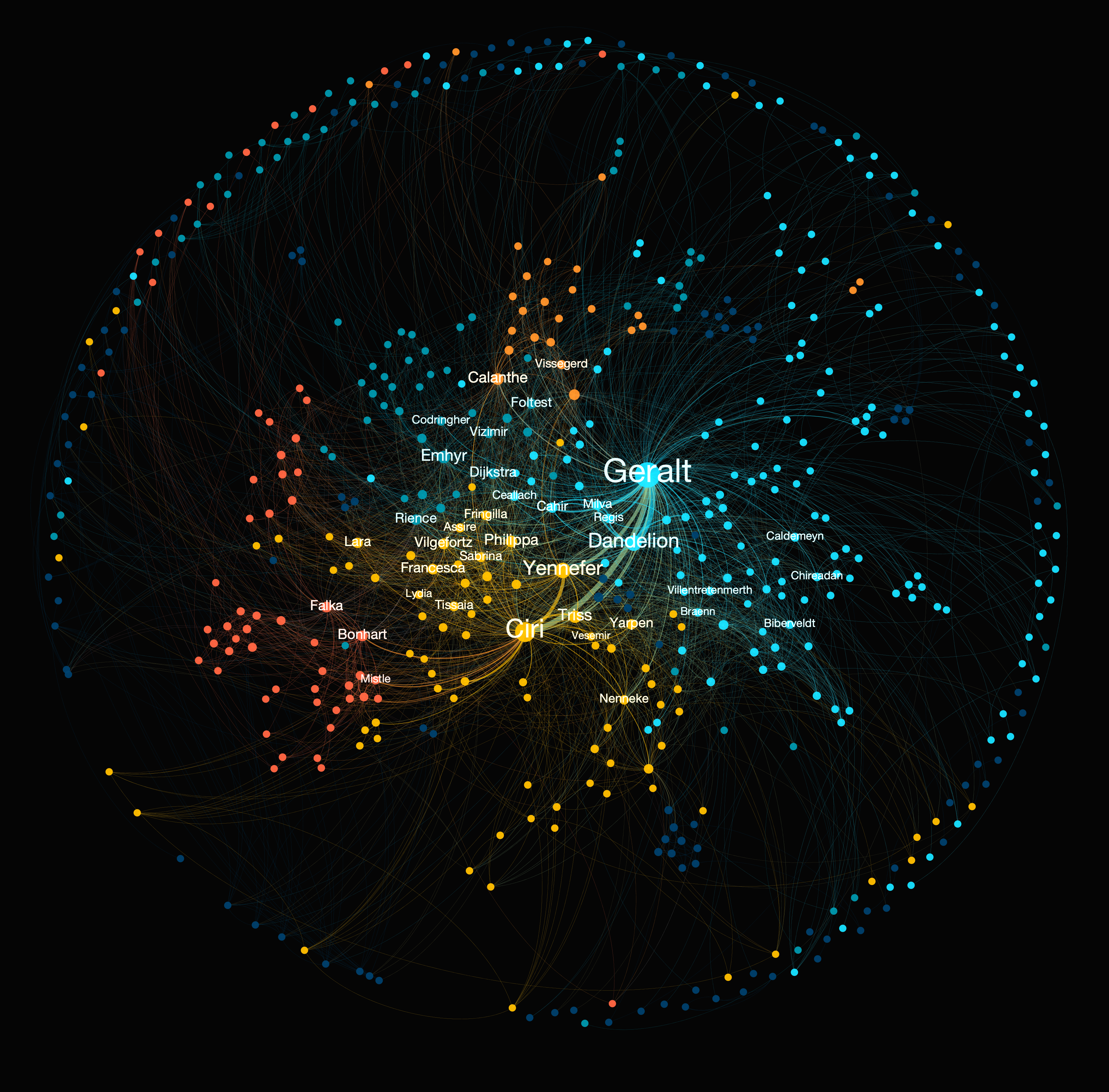}
\caption{The social map of {\it The Witcher}. Characters are represented by nodes, their size corresponding to their degree centrality, and their colors encode the network communities to which they belong. The network links are proportional to the number of times two characters were mentioned within a five-sentence distance from each other in the novel. The most significant 50 individuals are labeled}
\label{fig:fig3}
\end{figure}

\paragraph{Network communities.} 

Just as in real life, communities tell stories in {\it The Witcher}, too. More precisely, network communities are groups of nodes (characters) that are more closely connected than their connections to the rest of the network. These communities usually reveal strong signals about the main structure of networks: the presence, function, and relations of groups and people within. Therefore, following a widely used algorithm~\cite{comm}, I extracted these communities from {\it The Witcher}’s network and colored the nodes accordingly (Figure \ref{fig:fig3}).

You can immediately spot that this network contains two major communities: one centered by Ciri and the other centered by Geralt, the two main characters on their mission to find each other. Both communities touch each other at a number of points in the visualization. In addition, you can also see several peripheral communities bearing different colors.

On one hand, you can read from this graph that Geralt is surrounded by his travel companions Dandelion (Jaskier), Regis, Milva, and Cahir, who have been on a rescue mission to find Ciri — a storyline completely missing from the first two seasons of the TV show. On the other hand, Ciri has the strongest link to Yennefer and Triss (besides Geralt). Geralt has numerous weak connections reflecting his ever-changing journey. From the side of the sorceresses, Ciri is also linked to the dense community of mages marked by Philippa and Vilgefortz. Vilgefortz, logically following his conspiracy, is closely linked to Rience, who is in a community with Dijkstra and Codringher, being responsible for information and intelligence. Interestingly, Emhyr is also a part of this community, most likely due to him being mentioned many times in these conspiracies and the war between the North and Nilfgaard. Additionally, Falka (also known as Ciri) shows up in a distinct community corresponding to her adventures with the Rats, while the upper community in orange, signed by Calanthe, matched the early stories and Ciri’s birth in the court of Cintra.

If I overlay this picture with Figure \ref{fig:fig2}, you can immediately spot numerous differences — potential storylines for the future seasons to explore. For instance, in the case of Ciri, the entire saga of Ciri as Falka and her arch-enemy, Bonhart, is missing. And, Vilgefortz’s hideous plan is still unknown. You won’t find Geralt’s rescue mission in the Netflix show either, while Ciri’s training in the Temple of Melitele with Nenneke also appears with a much weaker community around it. Additionally, Fringilla being closely embedded into the network of mages highlights her completely different role and status in the novels versus the TV show–similar to the position of the elf leader, Francesca. Finally, the controversial Cahir has basically no ties to Geralt in the show – quite an opposite position from the one he holds in the novels.

\section{Summary}

In this article I demonstrated a way of turning characters from {\it The Witcher}’s novel series into a social network by mapping out nearby mentions of the hundreds of characters. Then I applied the toolsets of network science, a research field that examines complexity and connections, to uncover hidden information in this system of highly interconnected characters. This information included the influencers, the hubs, and the key characters and I related the cast of the novels to those from the currently aired TV show. Finally, I extracted and discussed the network’s community structure. From these analyses, you can conclude that a majority of the important characters from the novels have already been introduced on the show, while a number of communities are still missing. This implies that on one hand, following seasons will probably elaborate on the already running threads and on character development, adding a few spin-off plots, and probably showing some pronounced differences from the books, rewiring the novel’s network in unforeseeable ways.


\end{document}